# Heat transport and self-organized criticality in liquid $^4$He close to $T_\lambda$

**R. Haussmann**

*Sektion Physik der Ludwig-Maximilians-Universität München,
Theresienstrasse 37, D-80333 München, Germany*

*We present a renormalization-group calculation based on model F for the thermal conductivity $\lambda_{\rm T}(T,Q)$ in the presence of a homogeneous heat current $Q$ and gravity. For temperatures below $T_\lambda$ we obtain a large but finite thermal conductivity corresponding to superfluid $^4$He with dissipation. Furthermore, we consider the self-organized critical state where the effects of gravity and heat current cancel each other so that the distance from criticality $\Delta T = T(z) - T_\lambda(z)$ is constant in space and a function of $Q$. We compare our theoretical results with recent experiments.*

Liquid $^4$He at temperatures $T$ near the superfluid transition $T_\lambda$ in the presence of a homogeneous heat current $Q$ offers the possibility to study dynamical critical phenomena under nonequilibrium conditions. Theoretical[1,2] and experimental[3] investigations in the critical regime have shown that the superfluid transition is shifted to lower temperatures by a nonzero heat current $Q$. The experimentally observed critical temperature $T_{\rm c}(Q)$ is lower than the theoretically predicted $T_\lambda(Q)$, i.e. $T_{\rm c}(Q) < T_\lambda(Q) < T_\lambda$. While for $T \gtrsim T_\lambda(Q)$ the helium is normal fluid and for $T \leq T_{\rm c}(Q)$ it is superfluid, Liu and Ahlers[4] found a new dissipative region for temperatures in the interval $T_{\rm c}(Q) < T < T_\lambda(Q)$. This observation indicates that at a finite heat current $Q$ the transition from normal fluid to superfluid helium may happen in two steps with two critical temperatures $T_\lambda(Q)$ and $T_{\rm c}(Q)$. A similar experiment was performed by Murphy and Meyer.[5]

In this letter we present a renormalization-group (RG) theory for the thermal conductivity $\lambda_{\rm T} = \lambda_{\rm T}(T,Q)$ and the temperature profile $T(z)$ of

R. Haussmann

liquid $^4$He in the presence of a nonzero heat current $Q$ and gravity for all temperatures above and below $T_\lambda$. We intend to find an explanation of the experimental observations. In a previous RG theory[6] $\lambda_T(T,Q)$ has been calculated in the normal-fluid region for $T \gtrsim T_\lambda$ without gravity. Here we extend these calculations into the superfluid region $T \lesssim T_\lambda$. While in the previous theories[1,2] the dissipation of the superfluid-normal-fluid counterflow related to $Q$ was neglected and hence infinite thermal conductivities were obtained for $T < T_\lambda(Q)$, here we present a RG theory for the first time which *includes* the dissipation so that we obtain a large but finite $\lambda_T(T,Q)$ for $T < T_\lambda(Q)$.

The experiments[3-5] have been performed on earth under the influence of gravity. While in most cases gravity is negligible, for small heat currents below $1\,\mu\mathrm{W/cm}^2$ and temperatures closer than $0.1\,\mu$K to $T_\lambda$ gravity effects are observable and become important. Via the hydrostatic pressure variation induced by gravity the superfluid transition temperature $T_\lambda = T_\lambda(z)$ depends on the altitude coordinate $z$ where the gradient is $\partial T_\lambda/\partial z = +1.273\mu$K/cm.[7] In the experiments[3-5] the heat current $Q$ flows vertically from bottom to top to avoid convection. The local thermodynamic state of the helium depends on the temperature difference $\Delta T(z) = T(z) - T_\lambda(z)$ and on $Q$. Since here the gradients $\nabla T$ and $\nabla T_\lambda$ have opposite signs, the effects of gravity and heat current are additive.

If the heat current flows downwards from top to bottom then the gradients $\nabla T$ and $\nabla T_\lambda$ have the same sign. The effects of gravity and heat current may cancel each other so that the difference $\Delta T = T(z) - T_\lambda(z)$ is constant in space. This configuration has been considered by Onuki[8] and suggested for an experiment by Ahlers and Liu.[9] A self-organized critical state may form which is homogeneous over a large region in space. The experiment has been realized recently by Moeur et al.[10] The constant temperature difference $\Delta T = \Delta T(Q)$ has been measured as a function of the heat current $Q$. We calculate $\Delta T(Q)$ within our theory and compare with the experiment.

In liquid $^4$He close to the superfluid transition heat-transport phenomena are described by model $F$[11] which is given by the Langevin equations for the order parameter $\psi(\mathbf{r},t)$ and the entropy variable $m(\mathbf{r},t)$:

$$\frac{\partial \psi}{\partial t} = -2\Gamma_0 \frac{\delta H}{\delta \psi^*} + ig_0 \psi \frac{\delta H}{\delta m} + \theta_\psi , \qquad (1)$$

$$\frac{\partial m}{\partial t} = \lambda_0 \nabla^2 \frac{\delta H}{\delta m} - 2g_0 \mathrm{Im}\left(\psi^* \frac{\delta H}{\delta \psi^*}\right) + \theta_m , \qquad (2)$$

where

$$H = \int d^d r [\tfrac{1}{2}\tau_0(z)|\psi|^2 + \tfrac{1}{2}|\nabla\psi|^2 + \tilde{u}_0|\psi|^4$$

**Heat transport and self-organized criticality in liquid $^4$He**

$$+\tfrac{1}{2}\chi_0^{-1} m^2 + \gamma_0 m|\psi|^2 - h_0 m] \tag{3}$$

is the free energy functional and $\theta_\psi$ and $\theta_m$ are Gaussian stochastic forces which incorporate the fluctuations. The heat current $Q$ is imposed by boundary conditions. The gravity is included via the temperature parameter $\tau_0(z)$ in (3) which is related to $T_\lambda(z)$ and depends linearly on the altitude $z$. Usually, the model is treated by field theoretic means. Critical fluctuations close to $T_\lambda$ are taken into account by renormalization and application of the RG theory. All the renormalized coupling parameters of model $F$ depending on a RG flow parameter and the nonuniversal amplitudes are known[12] so that model $F$ can be used for explicit calculations of physical quantities in the critical regime without any adjustable parameters.

The previous field-theoretic approach[2,6] is constructed as a perturbation theory in terms of Feynman diagrams starting with the mean-field solutions $\psi_{\mathrm{mf}}(z)$ and $m_{\mathrm{mf}}(z)$ of (1) and (2). The quantities $n_{\mathrm{s}}(z) = \langle |\psi|^2 \rangle$ and $\mathbf{J}_{\mathrm{s}}(z) = \langle \mathrm{Im}[\psi^* \nabla \psi] \rangle$ must be calculated perturbatively so that eventually $T(z)$, $\lambda_{\mathrm{T}}(z)$, and hence $\lambda_{\mathrm{T}}(T,Q)$ can be determined. Below $T_\lambda(Q)$ in the superfluid region the ansatz $\psi_{\mathrm{mf}}(z) = \eta\, e^{ikz}$ and $m_{\mathrm{mf}}(z) = \mathrm{const.}$ has been used[2] which, however, implies a constant temperature profile. Consequently, the gradient $\nabla T$ is zero and the thermal conductivity $\lambda_{\mathrm{T}}$ is infinite so that dissipation of the superfluid-normal-fluid counterflow is not included. On the other hand, in the superfluid state dissipation occurs by creation of vortices which implies strong fluctuations of the phase of the complex field $\psi$ so that $\langle \psi \rangle = 0$. This fact would require the ansatz $\psi_{\mathrm{mf}}(z) = 0$ which implies that $m_{\mathrm{mf}}(z)$ is a linear function of $z$. By using this ansatz the thermal conductivity $\lambda_{\mathrm{T}}(T,Q)$ was calculated previously[6] for nonzero $Q$ in the normal-fluid region where $T \gtrsim T_\lambda$. Since $n_{\mathrm{s}}(z)$ and $\mathbf{J}_{\mathrm{s}}(z)$ were evaluated by using the free $\psi$-field Green's function, this approach is is invalid below $T_\lambda$. Thus, dissipation in superfluid helium is not accessible by perturbation theory starting with a mean-field solution of the model-$F$ equations.

To overcome this problem we suggest the following idea and approximation. We generalize model $F$ by replacing the complex field $\psi$ by a vector $\Psi = (\psi_1, \ldots, \psi_n)$ of $n$ complex fields. While model $F$ is recovered for $n = 1$, here we consider the limit $n \to \infty$. It is well known that models involving Ginzburg-Landau functionals like the $\phi^4$ model can be solved exactly in this limit (see e.g. Ref. 13). The same is true also for model $F$. Thus, in the limit $n \to \infty$ the complete temperature profile $T(z)$ and the thermal conductivity $\lambda_{\mathrm{T}}(T,Q)$ can be calculated exactly. Eventually we set $n = 1$, apply the renormalization-group theory to include critical fluctuations and use the result as an approximation.[14] However, first the Green's functions must be calculated. While for the $m$-field Green's function the free Green's function is obtained in the limit $n \to \infty$, the $\psi$-field Green's function is determined by the Dyson equation together with the self energy shown in Fig. 1. In the limit $n \to \infty$ only the tadpole diagram survives, where however the full propagator line is thick and identified by the exact $\psi$-field Green's function. The dashed line represents the interaction mediated by the parameter $\tilde{u}_0$ in (3) and the $m$-field Green's function. By solving the Dyson equation we obtain self-consistent equations for two effective $z$-dependent parameters related to the temperature profiles $T(z)$ and $T_\lambda(z)$. These equations involve $n_s(z)$ and $\mathbf{J}_s(z)$ as one-loop integrals which are evaluated approximately by using the formulas of Ref. 6.

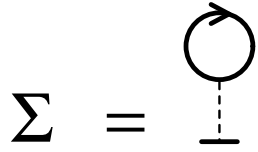

Fig. 1. The $\psi$-field self energy in the limit $n \to \infty$.

The calculations are very long and complicated and will be published elsewhere.[15] Eventually, we obtain the following results. The temperature difference and the thermal conductivity are given by

$$\Delta T(z) = T(z) - T_\lambda(z) = T_\lambda \, \tau \, \sigma^{1/3} \zeta \, (1 + 8u[\tau]A) \,, \tag{4}$$

$$\lambda_{\mathrm{T}}(z) = \frac{g_0 k_{\mathrm{B}}}{8\pi \, \xi(\tau)} \frac{(1 + \tfrac{1}{2} f[\tau] A_1)}{\tau \, \gamma[\tau] F[\tau]} \,, \tag{5}$$

respectively, where

$$A = -\frac{\sigma^{-1/6}}{2\pi^{1/2}} \zeta^{-1} \mathcal{F}_{-1/2}(\zeta) - 1 \,, \tag{6}$$

$$A_1 = \frac{\sigma^{-1/6}}{2\pi^{1/2}} \mathcal{F}_{1/2}(\zeta) - 1 \,, \tag{7}$$

$$\sigma = -\frac{1}{12} \Big[ (\rho_1')^2 + 2 \frac{w''[\tau]}{w'[\tau]} \rho_1' \Big( \frac{F[\tau] \Delta\rho'}{4\gamma[\tau] w'[\tau]} \Big) - \Big( \frac{F[\tau] \Delta\rho'}{4\gamma[\tau] w'[\tau]} \Big)^2 \Big] \,, \tag{8}$$

$$\Delta\rho' = -\frac{8\pi \, \gamma[\tau] F[\tau]}{1 + \tfrac{1}{2} f[\tau] A_1} \Big( \frac{Q[\xi(\tau)]^2}{g_0 k_{\mathrm{B}} T_\lambda} \Big) \,, \tag{9}$$

$$\rho_1' = \Big[ \Delta\rho' - \frac{1}{\tau} \frac{\xi(\tau)}{T_\lambda} \frac{\partial T_\lambda(z)}{\partial z} \Big] \Big/ (1 + 8u[\tau] A_1) \,, \tag{10}$$

$$\sigma^{1/3} \left[ 8 + \zeta - 16 u[\tau] A \, \zeta \right] = 1 \,. \tag{11}$$

Eqs. (6)-(11) are six equations for seven unknown variables $\zeta$, $\sigma$, $\tau$, $A$, $A_1$, $\Delta\rho'$, $\rho_1'$. Thus, there will be one parameter which can be varied arbitrarily.

**Heat transport and self-organized criticality in liquid $^4$He**

Preferably we choose $\zeta$ for this purpose and determine the other six variables by solving the equations. Here $u[\tau]$, $\gamma[\tau]$, $w'[\tau]$, $w''[\tau]$, $F[\tau]$, and $f[\tau]$ are the dimensionless renormalized couplings of model $F$ which are known functions depending on the renormalization-group flow parameter $\tau$.[12] Furthermore we need the correlation length $\xi(\tau) = \xi_0 \tau^{-\nu}$ with $\nu = 0.671$ and $\xi_0 = 1.45 \times 10^{-8}$ cm and the parameter $g_0 = 2.164 \times 10^{11}\,\text{s}^{-1}$ which is related to the entropy density at $T_\lambda$.[16,17] The amplitudes $A$ and $A_1$ represent the renormalized one-loop contributions by $n_s$ and $\mathbf{J}_s$, respectively. The one-loop integrals are represented by the function $\mathcal{F}_p(\zeta)$ which is defined by the integral

$$\mathcal{F}_p(\zeta) = \int_0^\infty dv\, v^{p-1} \exp(-v^3 - v\zeta)\ . \tag{12}$$

This function is known from the previous renormalization-group theory[6] and is derived with the assumption that the curvature of the temperature profile $T(z)$ can be neglected on the scale of a few correlation lengths $\xi(\tau)$. In a renormalization-group theory there is one equation needed which fixes the flow parameter $\tau$. Here this equation is given by (11). While the flow-parameter equation may be arbitrary in general, here (11) is an optimal choice. For $\zeta \gg +1$ and $\zeta \ll -1$ we recover the standard flow-parameter equations of the equilibrium theory[12] for $T$ above and below $T_\lambda$.

In (4)-(11) the space coordinate $z$ does not appear explicitly. Instead we have $\zeta$ as the variable. After eliminating $\zeta$ from (4) and (5) we obtain the thermal conductivity $\lambda_T = \lambda_T(\Delta T, Q)$ for given heat current $Q$ as a function of the temperature difference $\Delta T = T(z) - T_\lambda(z)$. Eventually, the temperature profile $T(z)$ is obtained by solving the heat-transport equation $\mathbf{Q} = -\lambda_T \nabla T$ as a differential equation.

Now, by solving the equations (4)-(11) numerically we obtain the following results. The thermal conductivity $\lambda_T(\Delta T, Q)$ is shown in Fig. 2 in a semi-logarithmic plot versus $\Delta T$ for heat current $Q = 100$ nW/cm$^2$. The solid line is our theoretical prediction while the crosses represent the experimental data of Day et al.[18] While the equilibrium superfluid transition is located at $\Delta T = 0$, the depressed critical temperature $\Delta T_\lambda(Q) = T_\lambda(Q) - T_\lambda$ is indicated by the arrow in Fig. 2 and located close to the position where the solid line has its maximum slope. Clearly, the agreement between our theory and the experiment[18] is quite good for $\Delta T \gtrsim \Delta T_\lambda(Q)$. We note that our theory does not have any adjustable parameters. However, for $\Delta T \lesssim \Delta T_\lambda(Q)$ the thermal resistivity is very small so that here experimental data are not available with sufficient accuracy on a logarithmic scale. Thus, for these low temperatures our theoretical curve is a prediction. While the effects of gravity are included in (10) by the gradient of $T_\lambda(z)$ and the heat current is very small, it turns out that the solid line in Fig. 2 is only slightly influenced by gravity which can be neglected. We find qualitatively similar curves for the thermal conductivity $\lambda_T(\Delta T, Q)$ like this one shown in Fig. 2 also for other heat currents $Q$, where gravity effects are negligible for $Q \gtrsim 100$ nW/cm$^2$.

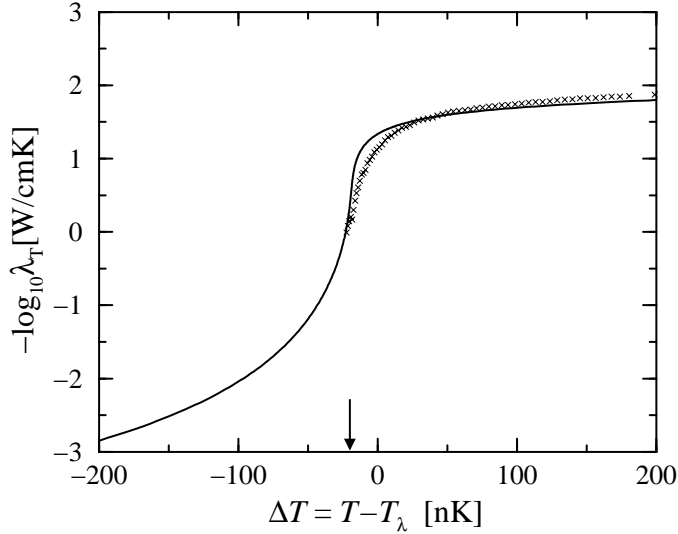

Fig. 2. The inverse thermal conductivity $\lambda_T^{-1}(\Delta T, Q)$ (thermal resistivity) as a function of $\Delta T(z) = T(z) - T_\lambda(z)$ for heat current $Q = 100$ nW/cm$^2$. The solid line is the prediction of our theory (4)-(12). The crosses represent the experimental data of Day et al.[18] where $Q$ flows upwards. The arrow indicates $\Delta T_\lambda(Q)$.

Liu and Ahlers[4] obtained the thermal conductivity $\lambda_T(\Delta T, Q)$ indirectly for $Q = 42.9$ $\mu$W/cm$^2$, which is shown in Fig. 1 of Ref. 4. Again for $\Delta T \gtrsim \Delta T_\lambda(Q)$ our theoretical prediction agrees with this result, while for $\Delta T \lesssim \Delta T_\lambda(Q)$ it does not agree. In the dissipative region $\Delta T_c(Q) < \Delta T < \Delta T_\lambda(Q)$ a considerably *smaller* thermal conductivity is observed than our theory predicts. The discrepancy may possibly be an artifact of the experiment caused by a surface effect, because in the experiment[4] endplate thermometers were used. In a more recent experiment,[19] where only sidewall thermometers were used, the dissipative region was not found. A thermal conductivity $\lambda_T$ was measured for $\Delta T < \Delta T_\lambda(Q)$ which qualitatively has the same temperature dependence as our theory but is about 20 times *larger*. Thus, a discrepancy remains, but now the opposite way round. Nevertheless, our present theory constitutes a progress and an extension of the previous theory to lower temperatures.

**Heat transport and self-organized criticality in liquid $^4$He**

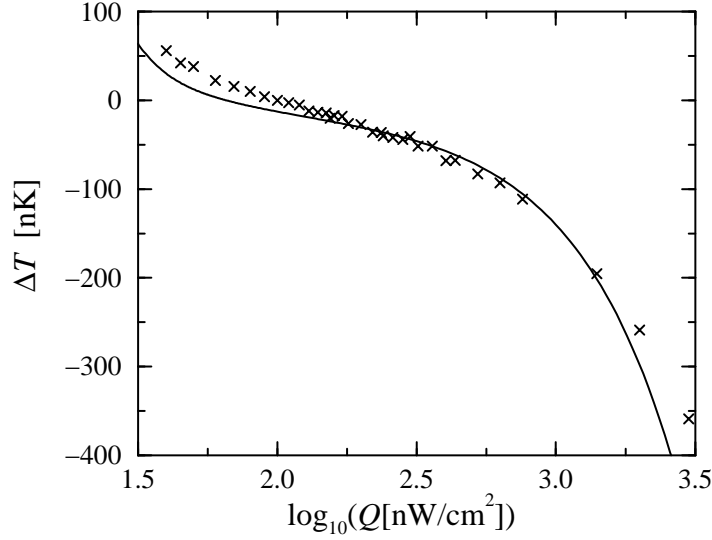

Fig. 3. The temperature difference $\Delta T = T(z) - T_\lambda(z)$ for the self organized critical state where the heat current flows downwards. The solid line is the prediction of our theory (4)-(12). The crosses represent the experimental data of Moeur et al.[10]

The self-organized critical state is obtained if the heat current flows downwards where $Q$ is negative and the temperature difference $\Delta T(z) = \Delta T$ is constant. Thus, the gradient of $\Delta T(z)$ is zero which implies $\rho_1' = 0$ in (10). Solving (4) and (6)-(11) with the constraint $\rho_1' = 0$ and eliminating all the seven unknown parameters, we eventually obtain $\Delta T = \Delta T(Q)$ as a function of $Q$. The result is shown in Fig. 3. Again, the solid line is our theoretical prediction, while the crosses represent the experimental data of Moeur et al.[10] While there are no adjustable parameters, the agreement between the theory and the experiment is quantitatively very good for heat currents $Q \lesssim 1.5$ $\mu$W/cm$^2$. However, for larger heat currents the experimental data deviate from our theoretical prediction. For a given $\Delta T$ the larger observed $Q$ means that the experimental thermal conductivity $\lambda_T$ is larger than theoretically predicted.

The strength of the critical fluctuations is indicated by the size of the correlation length $\xi(\tau) = \xi_0 \tau^{-\nu}$. We find that $\xi(\tau)$ is maximum for $\Delta T = \Delta T_\lambda(Q)$. Furthermore, we have calculated the specific heat $C_Q(\Delta T, Q)$ at constant heat current. We find a strong and relatively sharp maximum at $\Delta T = \Delta T_\lambda(Q)$. While the previous theories[20,21] predict a singularity at $T_\lambda(Q)$ for the specific heat, here the anomaly is rounded due to the finite temperature gradient. On the other hand, our theory does not predict anything unusual in thermodynamic and transport quantities at the lower critical temperature $T_c(Q)$ observed in the experiments.[3,5,10]

Furthermore, we have calculated the temperature profile $T(z)$ of the superfluid-normal-fluid interface. For heat currents $Q \gtrsim 1$ $\mu$W/cm$^2$ we expect that the effects of gravity are small. In the normal-fluid region we find that for these heat currents gravity is indeed negligible. However, in the superfluid region gravity effects are important even for larger heat currents. In microgravity we predict that $T(z)$ decreases monotonically without a lower bound for increasing $z$ which is due to the fact that the thermal conductivity $\lambda_T$ remains finite. In gravity where the heat current flows upwards we find a temperature $T(z) = T_c(Q)$ where the gradients $\nabla T(z)$ and $\nabla T_\lambda(z)$ have equal magnitude but opposite signs. For $T_c(Q) < T(z) < T_\lambda(Q)$ the gradient $\nabla T(z)$ dominates and in this region the temperature $T(z)$ decreases monotonically as in microgravity. However, for $T(z) < T_c(Q)$ the gradient $\nabla T_\lambda(z)$ dominates and for increasing $z$ the temperature approaches the limiting value $\lim_{z\to\infty} T(z) = T_\infty(Q)$ which is not much below $T_c(Q)$. Thus, our theory indeed yields a dissipative region and a superfluid region nearly without dissipation separated by a temperature $T_c(Q)$. The obtained shift $\Delta T_c(Q)$ is nearly the same as $\Delta T(Q)$ in the self-organized critical state, which has been found also in the experiments.[4,9,10] Our theory predicts that $T_c(Q)$ is not a property of liquid $^4$He but it is an artifact which is generated by gravity and occurs when integrating the heat-transport equation to obtain the temperature profile. In microgravity $T_c(Q)$ will not be present.

From these observations we conclude that our theory concerning $T_c(Q)$ agrees qualitatively with the experimental findings.[3–5,9,10] However, there are two major quantitative disagreements. First of all we predict $\Delta T_c(Q) \sim -Q^{1.25}$ for $Q \gtrsim 1$ $\mu$W/cm$^2$ which disagrees with the experimentally observed[3] $\Delta T_{c,\exp}(Q) \sim -Q^{0.81}$. Secondly, in the dissipative region $T_c(Q) < T(z) < T_\lambda(Q)$ the thermal conductivity $\lambda_T$ is much larger than found in the experiment[4] which implies that the spatial extent $\Delta z$ of the dissipative region is much larger. We find $\Delta z > 0.25$ cm for $Q > 10$ $\mu$W/cm$^2$ which is about the sample size in the experiments.[3–5] However, it is unclear if the dissipative regions and the related definitions of $T_c(Q)$ in theory and in the experiments have the same origin. The region of enhanced dissipation may possibly be a surface effect and not a property of the bulk helium, because it was not observed in the recent experiment[19] using only sidewall thermometers. Thus, it is unclear if the discrepancies are due to the theory or due to the experiment.

For the understanding of the nature of $T_c(Q)$ and of the dissipative region further experimental and theoretical investigations are necessary. The

**Heat transport and self-organized criticality in liquid $^4$He**

following questions should be addressed in future experiments. Does $T_c(Q)$ depend on gravity? Is $T_c(Q)$ absent in microgravity as our theory predicts? How large is the spatial extent $\Delta z$ of the dissipative region? (For this purpose, the spatial dependence of $T(z)$ should be investigated carefully.) Do thermodynamic quantities show any anomalies at $T_c(Q)$ or not? On the other hand, the discrepancies may be caused by the crucial assumption of our theory: we have generalized model $F$ by introducing an $n$-component complex order parameter and obtained our approximation by taking the limit $n \to \infty$. It turns out that for $n \geq 2$ already in mean-field approximation the superfluid-normal-fluid counterflow is unstable, while for $n = 1$ (model $F$) it is metastable. This fact may cause a qualitative difference between model $F$ and our approximation.

Nevertheless, our approach appears to describe the essential part of the physics correctly: we find dissipation for the superfluid-normal-fluid counterflow and obtain a finite thermal conductivity $\lambda_T$ in the superfluid region for nonzero heat currents $Q$. Since dissipation in the superfluid state is necessarily connected with the creation of vortices, our approximation must include vortices in an indirect way. We have applied our approach also to rotating helium.[22] In this case, the number of vortices is precisely known and related to the rotation frequency, so that the dissipative effect of a single vortex line can be extracted which is related to the Vinen coefficient[23] $B$. We obtain the simple result[22] $B = (4m_4/\hbar)\Gamma'[\tau]$ where $\Gamma'[\tau]$ is a renormalized coupling of model $F$ depending on the RG flow parameter $\tau$. This result has a similar temperature dependence and the same order of magnitude as the Vinen coefficient $B$ calculated within the renormalized mean-field theory by considering a single vortex line.[24] Thus, the large discrepancies of the thermal conductivities by about a factor of 20, which we have discussed above, are not caused by an incorrect treatment of the dissipative effect of the vortex lines. Rather, the discrepancies must be due to the density of the vortices in the helium, which in the homogeneous heat flow appears to be a very sensity quantity possibly influenced strongly by the kind of the approximation and by the conditions of the experiment.

Finally, our approach can be tested in thermal equilibrium for $Q = 0$ by calculating e.g. the specific heat $C_P$. Our result for $C_P$ agrees with the results of the previous theories[12] within the related accuracies. While the correct critical behavior is incorporated by the RG theory, we obtain the amplitude function $F[u]$ of the specific heat, which below $T_\lambda$ agrees with the amplitude function $F_-[u]$ of Ref. 12 in renormalized mean-field theory (zero-loop approximation) and above $T_\lambda$ agrees with $F_+[u]$ of Ref. 12 in one-loop approximation. Thus, we conclude that our approach is valid at least in thermal equilibrium.

R. Haussmann

I would like to thank Professors V. Dohm, R. V. Duncan, H. Meyer, and Dr. D. Murphy for discussions and for comments on the manuscript.